\documentclass[journal, 10pt]{IEEEtran}
\usepackage{graphicx}
\usepackage{caption}
\usepackage{subcaption}
\usepackage{multirow}
\usepackage{url}
\usepackage{array}
\usepackage{cite}
\usepackage{color}
\usepackage[table]{xcolor}
\usepackage{arydshln}
\usepackage{bbding}
\usepackage{pifont}
\usepackage{wasysym}
\usepackage{amssymb}
\usepackage{color,soul}
\usepackage{ulem}
\usepackage{comment}

\definecolor{rowcolor1}{RGB}{235, 235, 235} % Light grey
\definecolor{rowcolor2}{RGB}{220, 220, 220} % Gainsboro

\newcommand{\cmark}{\ding{51}}%

\newcommand{\papertitle}{Operationalizing AI/ML in  Future Networks: \\ A Bird's Eye View from the System Perspective}

\begin{document}

\title{\papertitle}

\author{Qiong~Liu,
        Tianzhu~Zhang,
        Masoud~Hemmatpour,
        Han~Qiu,
        Dong~Zhang,
        Chung~Shue~Chen,
        Marco~Mellia,
        Armen~Aghasaryan
%\thanks{Q. Liu is with the Department of Computer Sciences and Networks, Telecom Paris, Palaiseau, France. (E-mail: qiong.liu@telecom-paris.fr)}
%\thanks{T. Zhang, C. S. Chen, and A. Aghasaryan are with Nokia Bell Labs, Massy, France. (E-mail: \{tianzhu.zhang, chung\_shue.chen, armen.aghasaryan\}@nokia-bell-labs.com)} % <-this % stops a space
%\thanks{M. Hemmatpour is with Simula Research Laboratory, Oslo, Norway. (E-mail: mashemat@simula.no)}% <-this % stops a space
%\thanks{H. Qiu is with the Institute for Network Sciences and Cyberspace, BNRist, Tsinghua University, Beijing, China. (E-mail: qiuhan@tsinghua.edu.cn)}
%\thanks{D. Zhang is with the College of Computer Science and Big Data, Fuzhou University, Fuzhou, China. (E-mail: zhangdong@fzu.edu.cn)}
%\thanks{M. Mellia is with the Department of Control and Computer Engineering, Politecnico di Torino, Turin, Italy. (E-mail: marco.mellia@polito.it)}
}

%\markboth{IEEE Communications Magazine}%
%{T. Zhang \MakeLowercase{\textit{et al.}}: \papertitle}

\maketitle

\begin{abstract}
Modern Artificial Intelligence (AI) technologies, led by Machine Learning (ML), have gained unprecedented momentum over the past decade. Following this wave of ``AI summer'', the network research community has also embraced AI/ML algorithms to address many problems related to network operations and management. However, compared to their counterparts in other domains, most ML-based solutions have yet to receive large-scale deployment due to insufficient maturity for production settings. 
This article concentrates on the practical issues of developing and operating ML-based solutions in real networks. Specifically, we enumerate the key factors hindering the integration of AI/ML in real networks and review existing solutions to uncover the missing considerations. Further, we highlight a promising direction, i.e., Machine Learning Operations (MLOps), that can close the gap.
We believe this paper spotlights the system-related considerations on implementing \& maintaining ML-based solutions and invigorate their full adoption in future networks.
\end{abstract}

\begin{IEEEkeywords}
AI/ML for Networking, Network Systems
\end{IEEEkeywords}

\section*{Introduction}

The last decade has witnessed a thorough evolution of the modern telco industry with the advent of network softwarization techniques, such as Software-Defined Networking (SDN) and Network Function Virtualization (NFV).
% The ongoing 5G rollout promises to deliver customized network services to billions of subscribers with ultra-high speed, ultra-high reliability, ultra-low latency, and ubiquitous connectivity~\cite{3gpp2024ai}. 
By transforming traditional hardware-centric networking components into software-based processes, SDN/NFV authorizes unprecedented flexibility, scalability, and efficiency~\cite{swamy2022homunculus,yao2022aquarius,zheng2022automating}. Despite these benefits, with the rapid expansion of telco infrastructure, the scale and dynamism of modern networks keep growing, and network management remains a daunting task~\cite{rossi2022landing}.

Meanwhile, AI/ML makes remarkable advancements and has attracted strategic attention across various business sectors. According to Gartner and MIT Sloan Management, AI has led to \$3.9T of annual business value and is deemed a strategic priority by 83\% of CEOs~\cite{databricks}. Inspired by these successes, network researchers are extensively exploring AI/ML for diverse tasks~\cite{iacoboaiea2022design,yao2022aquarius}. %such as resource management~\cite{iacoboaiea2022design,yao2022aquarius} and troubleshooting~\cite{swamy2022homunculus,yang2021quality}. 
These {\em ML-based solutions}, %\footnote{Note that we focus on the data-driven branch of AI techniques, frequently denoted as ML. By abuse of language, we will use the two acronyms interchangeably in this paper.}, 
i.e., applications, functions, and services, have demonstrated more promising outcomes than traditional fixed-policy approaches~\cite{rossi2022landing}.

Despite the enormous interest, the modern network's fast-paced evolution has made it impossible to construct and manage large corpses of networking data, which were crucial for AI's successful deployment in real systems.
%Despite the enormous interest, AI/ML is still immature for deployment in real networks. 
According to a recent report~\cite{ava}, $88\%$ of the telco industry's proof-of-concept AI/ML projects fail to reach live deployment. The major deterrent stems from inadequate ``system thinking''~\cite{sculley2015hidden}.
% as researchers are not always exposed to the complexities of production networks~\cite{sculley2015hidden}.
Based on our observation, existing AI/ML-based solutions have two fundamental disparities with real-network deployments: (i) {\em One-dimensional design:} ML solutions mainly aim to outperform prior solutions on specific performance metrics, especially accuracy, without vetting other network-/system-critical imperatives. For example, %as networks usually operate at very high speeds, it is arduous to obtain high-quality data and features~\cite{bronzino2021traffic};
as network operations get increasingly complex and intertwined, optimization becomes multi-metric and multi-dimensional~\cite{huyen2022designing}; (ii) {\em System discrepancy:} These solutions were mostly demonstrated in controlled environments and became costly to fit into real network systems with much higher scale, complexity, and dynamism. For instance, given the data-driven nature of ML-based solutions, fulfilling performance guarantees under sporadic data and environment drifts is non-trivial~\cite{yang2021quality}.
This ``reality gap" greatly hampers the integration and deployment of AI/ML in real networks. 

To make AI/ML an integral part of modern networks, there is a need for lightweight techniques capable of timely prioritizing and triggering model updates, which guarantee the deployed models remain fit for their task regardless of environment evolvement. Given these premises, this paper aspires to elucidate the practical challenges of integrating AI/ML into the future network landscape. Specifically, we present network-oriented AI/ML research and its gap with real networks. Then, we enumerate the practical considerations to actualize AI/ML in production-ready networks. Afterward, we prospect a promising direction — MLOps, which applies agile methodologies to combine software development (Dev) and IT operations (Ops), aimed at shortening the systems development lifecycle and providing continuous delivery with high software quality~\cite{rossi2022network}. We finally introduce two example use cases in network softwarization about continual performance prediction and abnormal detection, where we apply several of the abovementioned techniques.

% In the information technology (IT) industry, the development and delivery of software products are usually streamlined via Development and Operations (DevOps) practices, which are further customized by the telco industry to enhance service quality and reduce time-to-market~\cite{john2017service}.
%Although network-oriented DevOps practices offset part of the exertions via continuous integration and delivery, they can hardly accommodate the unique characteristics of AI/ML~\cite{yang2021quality}. To smoothly {\em operationalize} (i.e., {\em develop}, {\em deploy}, and {\em manage}) AI/ML in production, network operators must master skills in ML, data engineering, and systems, which is extremely burdensome. 

%We believe MLOps can relieve the associated operational overhead by adopting time-tested practices. At the same time, Causal AI navigates a novel path toward verifiable AI via casual reasoning and hypothesis testing. Both will play an important role in the future path to adopting AI/ML for networks.

%{\color{blue} Chip Huyen's book \cite{huyen2022designing}.}

%\section*{Background}\label{sec:background}
\section*{Landing AI in networks}\label{sec:background}
In this section, we briefly review the current status of AI/ML and elaborate on the practical barriers obstructing their general adoption in operational networks.

\iffalse
**************reviewer 2***************
\footnote{\textcolor{blue}{Reviewer 2: Section numbering is not within the Magazine's style, and the authors are advised to refer to prior or following sections in a non-numbered fashion.}}
**************reviewer 2***************
\fi

\subsection*{Current States}

In recent years, AI/ML has sparked tremendous hype in the operational networks thanks to (i) the innovative breakthroughs in theoretical research, (ii) the success in other fields such as computer vision and NLP, and (iii) the presence of optimized development toolkits with hardware acceleration. 
Compared to fixed-policy approaches, AI/ML algorithms exhibit exceptional pattern matching, incremental learning, and automation capabilities on large-scale, multidimensional data~\cite{iacoboaiea2022design}. 

Standardization bodies (e.g., ETSI, 3GPP) anticipate AI/ML techniques to be crucial in automating future networks. % and have formed multiple working groups to investigate different use cases.
%\textcolor[HTML]{006400}{
In February 2024, ETSI released a standard (ETSI TR104032 ~\cite{etsi_tr_104032}), which highlighted the necessity of logging key details throughout an AI model's lifecycle by using model trace records like the MLOps framework.
%Additionally, a 3GPP standard~\cite{3gpp2024ai} mentioned the importance of management capabilities and services to support and integrate AI/ML in 5Gs.
Moreover, a 3GPP standard (Rel-17\cite{3gpp2024ai}) underscored the necessity for management tools and services to facilitate the incorporation of AI/ML technologies in 5G networks. 
%}

\iffalse
********************reviewer 2********************

\footnote{\textcolor{blue}{reviewer2:as the authors note, ETSI and 3GPP have provided architectural descriptions of the position of AI/ML solutions. How does the use of MLOPs align with these?}}

********************reviewer 2********************
\fi
In industry, carrier-grade platforms are under active development to bolster AI/ML-augmented network services: Nokia's AVA Ecosystem offers telco operators cloud-native AI/ML and analytic services to automate network operations, enhance service assurance and subscriber experience and reduce cost~\cite{ava}; Huawei's ADN ecosystem features network automation with dedicated support for AI operations~\cite{rossi2022landing}, which consists of three tiers, i.e., on-device AI, online fog/cloud AI, and offline cloud AI, to support network and AI operations with assorted temporal-spatial properties.  
In academia, ML algorithms are widely developed to tackle a large spectrum of "networking" problems, such as traffic classification~\cite{yang2021quality}, resource scheduling~\cite{iacoboaiea2022design}, anomaly detection~\cite{swamy2022homunculus}, load balancing~\cite{yao2022aquarius}, QoE management~\cite{bronzino2021traffic}.
%\qiuhan{Consider using one of the above examples in the Intro to illustrate the challenges.} 
Given the rapid expansion of the AI/ML frontier (e.g., generative AI), their growth in telco networks will continue to enrich. However, there is still a certain distance between proof-of-concept and successful real-time deployment of AI/ML projects. We will discuss the specific difficulties in the following sections.

\subsection*{Challenges \& Barriers}
\begin{figure}
\begin{center}
\includegraphics[width=0.49\textwidth,height=0.125\textheight]{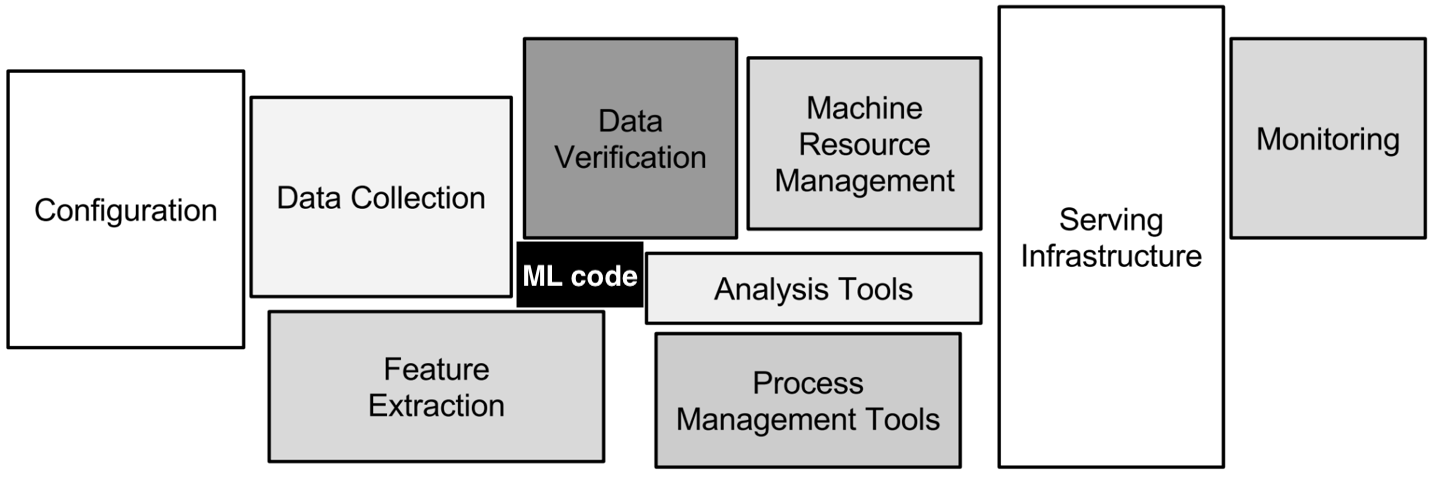}
\caption{Basic components for real-world ML systems (Picture originated from Sculley et al.~\cite{sculley2015hidden}).}
\label{fig:debts}
\end{center}
\end{figure}

The term ``ML system" is frequently associated with the algorithms it employs, like logistic regression or various neural networks. However, these algorithms only represent a fraction of a full ML system in a production environment. As mentioned in Fig.~\ref{fig:debts}, ML systems in the real world encompass the initial business objectives, the interfaces, the entire data stack, and the methodologies for model development, monitoring, and updating. 
% Additionally, it includes the underlying infrastructure that facilitates the execution and process delivery. 
ML in production does not connote ML in research, as the latter seldom bothered with the deployment and maintenance issues once the optimization goal was achieved on the test dataset~\cite{rossi2022landing}. Based on our study, the key challenges of landing AI in networks can be summarized as follows:
%}

\iffalse
********************reviewer 1******************
\footnote{\textcolor{red}{Reviewer 1: The authors mention at several places that ML/AI solutions are generally inadequate for live deployment in real network systems. The reality gap presented by the authors falls short to prove their claim.}}
********************reviewer 1******************
\fi

%\textcolor{blue}{Despite the plethora of solutions, a closer inspection reveals a less rosy picture. These solutions are generally inadequate for live deployment in real network systems~\cite{ava}. } %\footnote{\textcolor{blue}{reviewer2: The hyphen in ".. less rosy picture - These solutions ..." should be replaced.}} 

%Based on our study, such a reality gap originates from three factors:  
% Most were implemented in a highly empirical and manual fashion throughout the AI/ML lifecycle stages, including data acquisition, feature extraction, algorithm design, model training, parameter tuning, and validation. Also, they seldom bothered with the deployment and maintenance issues once the optimization goal was achieved on the test dataset.
%Such an approach is acceptable for fast prototyping yet unsustainable for production environments. 

\subsubsection*{\bf Data complexity}
%Compared to other prevalent AI/ML application domains, e.g., computer vision and language processing, 
Network data has much more diverse formats, e.g., raw packets, flow-level statistics, configuration files, system logs, and event alarms. They may contain categorical, temporal, spatial, or even graph semantics. Such multi-modal data with high variety, velocity, and volume can be exceedingly onerous to model and process~\cite{bronzino2021traffic}, not to mention their natural distribution drifts caused by data and system evolvements. 

\begin{figure*}[!tb]
    \centering
    \includegraphics[width=1\textwidth, height=0.12\textheight]{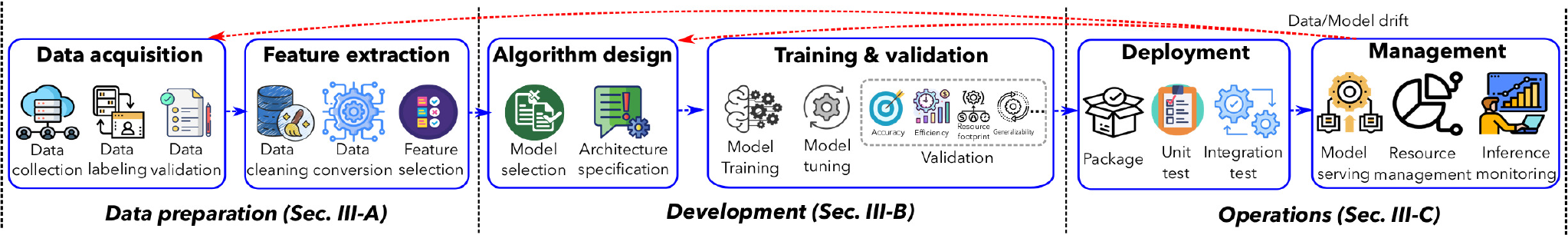}
    \caption{ML lifecycle in production settings.}
    \label{fig:pipeline}
\end{figure*}

\subsubsection*{\bf Multi-dimensional requirements nature}

Researchers often align on one single objective. The most common objective is model performance: developing a model that achieves state-of-the-art results on benchmark datasets.
% To edge out a small performance improvement, researchers often resort to techniques that make models too complex to be useful.
In production networks, KPI optimization cannot be done in isolation. For example, some DNN models with high prediction accuracy can hardly fit into resource-limited network devices~\cite{zheng2022automating}. Besides, the potential high inference latency can make the model unsuitable for real-time requirements, particularly in high-speed networks where the service latency is measured in microseconds~\cite{bronzino2021traffic}. In essence, the learning and run-time complexity in ML systems should be equivalently considered: the former pertains to the computational \& resource costs associated with developing an ML model, and the latter refers to the costs of deploying and managing a trained model.
%ML engineers need to ensure the multidimensional effectiveness of solutions, preventing one-dimensional solutions that optimize for a single performance metric.}

%Existing solutions generally focus on optimizing specific performance metrics rather than comprehensively assessing the overall readiness, which is incompatible with the stipulations of real-world ML systems. In particular, some solutions strive for high prediction accuracy using supersized Deep Neural Networks (DNNs) models that can hardly fit into resource-limited network devices. The potentially high inference latency can make them unsuitable for real-time constraints. They may also interfere with the critical data path. But most of all, in production networks, all Key Performance Indicators (KPIs) are naturally intertwined and must be jointly analyzed and attuned to network/system constraints to avoid one-dimensional solutions.

\subsubsection*{\bf Hidden technical debts}
This term was coined by Sculley et al.~\cite{sculley2015hidden}, which refers to the massive operational costs of operationalizing ML-based systems by non-experts. 
Similar debts also apply in network systems. As existing solutions were mostly developed in simulated or controlled environments, the practical deployment and maintenance issues were usually sidelined. In real systems, instead, ML models should be deployed as part of a data-processing pipeline. Owing to disparate development toolkits and deployment targets, integrating them into real networks can be laborious and error-prone. As network devices can come from sundry vendors with bespoke configuration, optimization, and execution routines, deploying AI/ML on them can result in complicated manual tuning, customization, and feasibility tests. In addition, rather than a one-off process, ML-based solutions must be continuously upgraded to meet business requirements and sustain long-term value over the rapid evolution of the telco industry. 

%Lastly, conventional ML algorithms possess limited generalization capability, as their success is largely built upon the independent and identically distributed (i.i.d) assumption~\cite{scholkopf2021toward}. If this assumption breaks, the accuracy of most ML models will only plummet. Natural or interventional data drift can happen in real networks, leading to training-serving skew and rendering a model obsolete. Furthermore, though conventional ML algorithms excel in making predictions within i.i.d settings and beyond, they still struggle with problems that demand causal or counterfactual reasoning~\cite{jakovljevic2021towards}. The crux of this limitation is the correlation-centric pattern matching, which often fails to capture the true causal relationships of the underlying data generation mechanism, e.g., the latent confounding factors can inject spurious correlations to bewilder an ML algorithm~\cite{scholkopf2021toward}.
%Therefore, before fully embracing AI/ML in production networks, a pragmatic question is whether their outcomes can always be explained, verified, and trusted.

\section*{Operationalizing AI/ML in production networks: the status quo} \label{sec:challenges}
%To close the gap and seamlessly operationalize AI/ML in production, many critical system-related considerations exist throughout the ML lifecycle, i.e., data preparation, development, and operations phases, as illustrated in Fig.~\ref{fig:pipeline}. This section epitomizes the relevant ones for operational networks.

%\ql{To ensure the seamless integration of AI/ML into production environments, it's imperative to address numerous pivotal considerations throughout the ML lifecycle depicted in Fig.~\ref{fig:pipeline}. This section encapsulates these considerations and explores associated studies within the networking domain. The included works were chosen based on their adherence to two criteria: (i) they address one or more practical aspects, and (ii) the methodologies or frameworks proposed have undergone implementation and verification within actual network systems.}

To close the gap and seamlessly operationalize AI/ML in production, many critical system-related considerations exist throughout the ML lifecycle, i.e., data preparation, development, and operations phases, as illustrated in Fig.~\ref{fig:pipeline}. This section encapsulates these considerations and explores associated studies within the networking domain. The included works were chosen based on two criteria: (i) they address one or more practical aspects, and (ii) the methodologies proposed have undergone implementation and verification within actual network systems.

\subsection*{Data preparation}~\label{sec:data}
\label{sec:data_pre} \vspace{-3mm}
%\QL{As the cornerstone of modern AI, data quality directly determines the ceilings of any AI/ML-based product, spurring the recent trend towards data-centric AI~\cite{databricks}. However, ensuring data quality can be extremely time-consuming, often costing $60\%$ % of the time in AI/ML projects~\cite{ava}. 
%To supply the ML algorithms with high-quality data, special considerations should be enforced in the data preparation phase: the constituent {\em data acquisition} and {\em feature extraction} processes. }

Data quality directly determines the ceilings of any AI/ML-based product, spurring the recent trend towards data-centric AI~\cite{databricks}. Due to the complexities in real networks, good datasets are not always available. Ensuring data quality can averagely cost $60\%$ of time in AI/ML projects~\cite{ava}. 
Special considerations should be enforced upon data preparation to supply the ML algorithms with high-quality data:  the constituent {\em data acquisition} and {\em feature extraction} processes.

\textbf{Data acquisition.}
As supervised learning is the most applied algorithm, obtaining labels is integral to creating training data~\cite{huyen2022designing}.
In existing solutions, data can generally originate from three sources: (i) live networks, (ii) controlled environments, or (iii) (curated) public data/datasets. 
%\textcolor[HTML]{006400}{%We make a distinction between ``data" and ``dataset." The term ``dataset" refers to a collection that is both finite and static. Conversely, data encountered in production is neither limited in quantity nor fixed in nature, and it also experiences distribution shifts.}
In case (i), despite the various data collection methods, the process can incur huge operational costs, which obligates considerate tradeoffs~\cite{yao2022aquarius}. For example, sampling is usually prioritized over per-packet collection in high-speed networks to attenuate the impact on the datapath. Also, data collection can incur uncontrollable situations, such as packet drops, sampling biases, or schema changes, hence aberrations and outliers. %\textcolor[HTML]{006400}{Besides, despite the promise of unsupervised ML and reinforcement learning, most ML models in production today are supervised, which means that they need labeled data to learn from.} 
Data labeling remains laborious as it consumes substantial human effort and does not scale with data volume~\cite{yang2021quality}.
Despite the advanced techniques (e.g., weak supervision, semi-supervision, transfer learning, and active learning) to mitigate data scarcity, these methods still depend on pre-labeled datasets or human input, limiting their scalability and efficacy in handling large, complex datasets.
In cases (ii) and (iii), as data are from outside the target networks, its statistical properties can be unaligned with deployment assumptions, leading to unexpected consequences, such as data drifts.
%This misalignment can trigger model performance degradation due to data drift, where the model is exposed to operational data whose distribution deviates from the training data. 
Thus, testing becomes necessary to disclose potential biases/anomalies before model deployment.

\textbf{Feature extraction.} Raw network data must be converted to features conformant with the ensuing AI/ML algorithms. Feature extraction is challenging - different feature sets imply varied system costs (and model performance), thus merit closer scrutiny: many existing ML-based solutions empirically define custom features, which may become hard to obtain and scale in deployment. Furthermore, feature selection schemes, when applied, might face revamping upon network evolution. As detailed in \cite{holland2021new}, traffic patterns and network conditions in real systems always shift, rendering existing features obsolete and necessitating engineering new features. 

%{\color{red} An example of a potential complication is data leakage during feature extraction. This occurs when information too closely related to the target variable is utilized, resulting in deceptively high-performance metrics during the training phase but poor live deployment performance.}
%\textcolor[HTML]{006400}{For example, data leakage during feature extraction may happen when information too closely tied to the target variable is used, leading to misleadingly high performance during training but poor real-world application~\cite{holland2021new}.}

%In essence, both data acquisition and feature extraction should be carefully designed to prepare the most relevant features for model development.

\textit{Existing solutions:} In contemporary network research, several seminal works approached the practical challenges of data acquisition and feature extraction: Bronzino et al.~\cite{bronzino2021traffic} introduced Traffic Refinery, an efficient automation pipeline for flow-level data collection and feature extraction. It aligns network operator goals by consolidating multiple design choices to alleviate packet losses. Additionally, a dedicated profiler quantifies system-level costs, offering operators a trade-off between feature selection and model accuracy. In a distinct exploration, {Yao et al.~\cite{yao2022aquarius}} proposed the {Aquarius} framework to enable flexible data collection and feature extraction for data center networks. This system embeds a transport-layer collector for effective TCP traffic feature extraction, storing them in shared memory to facilitate seamless ML algorithm interactions on the control plane, devoid of data plane disruption. Lastly, {Holland et al.~\cite{holland2021new}} proposed the {nPrint} framework, which transforms packets into a consistent binary format without sacrificing contextual meaning. This mechanism empowers ML algorithms to automatically identify key features, avoiding the efforts of manual feature extraction. %Their {\em nPrint} prototype boasts an impressive encoding rate, handing up to $10^6$ packets/minutes.

%In summary, these solutions represent a broader trend in network research toward developing more sophisticated, ML-compatible methods for data preparation. Ongoing work is still required to ensure that data acquisition and feature extraction are practical, secure, and beneficial in real networks.

%Overall, data preparation involves direct interaction with real networks, which can be extremely challenging given the high-speed data flows and system-level complexities. Despite the limited test scenarios, these solutions provide valuable first-hand guidelines for efficiently collecting \& representing data and initiate a broader trend toward developing more sophisticated, ML-compatible methods for data preparation. 

\begin{table*}[!tb]
\caption{Synoptic of the related works.}
\setlength{\tabcolsep}{1pt}
\renewcommand{\arraystretch}{0.7}
%\scriptsize
\begin{center}
\begin{tabular}{|c|cccccccc|c|c|c|}
\hline 
 \rowcolor{rowcolor2}
& \it Data & \it Feature & \it Algorithm & \it Hyperparam. & \it Model &  & & & {\bf Target} & \\
 \rowcolor{rowcolor2}
\multirow{-2}{*}{\cellcolor{rowcolor2}\bf Reference}  & \it acquisition & \it extraction &\it design &\it tuning & \it training & \multirow{-2}{*}{\it Validation} & \multirow{-2}{*}{\it Deployment} & \multirow{-2}{*}{\it Management}  & \bf network & \multirow{-2}{*}{\bf Use cases}  \\ \hline
 \hline
 \rowcolor{rowcolor1}
 {\bf Bronzino et al.~\cite{bronzino2021traffic}}  & \cmark & \cmark &&&&&& & - & QoE inference \\ \hline
 \multirow{3}{*}{\bf Yao et al.~\cite{yao2022aquarius}} & \multirow{3}{*}{\cmark} & \multirow{3}{*}{\cmark}  &&&&&& & \multirow{2}{*}{Datacenter} & Load balancing \\
&  &&&&&&& &  \multirow{2}{*}{network} & Traffic classification \\ 
&  &&&&&&&  && Resource scheduling \\\hline
\rowcolor{rowcolor1}
{\bf Holland et al.~\cite{holland2021new}} & & \cmark & \cmark & \cmark & \cmark & \cmark && & - & Traffic analysis \\\hline
\multirow{3}{*}{\bf Swamy et al.~\cite{swamy2022homunculus}} & & & \multirow{3}{*}{\cmark} & \multirow{3}{*}{\cmark} & \multirow{3}{*}{\cmark} &  \multirow{3}{*}{\cmark} & \multirow{3}{*}{\cmark} & & \multirow{2}{*}{Datacenter}  & Anomaly detection \\
&&&&&&&  && \multirow{2}{*}{network}  & Traffic classification \\
&&&&&& & & && Botnet detection \\\hline
\rowcolor{rowcolor1}
{\bf Lacoboaiea et al.~\cite{iacoboaiea2022design}} &&&&& \cmark & \cmark && & WLAN & Resource scheduling \\ \hline
\multirow{2}{*}{\bf Zheng et al.~\cite{zheng2022automating}} &&&&& \multirow{2}{*}{\cmark}  & \multirow{2}{*}{\cmark} & \multirow{2}{*}{\cmark} & \multirow{2}{*}{\cmark} & Datacenter & Anomaly detection \\
&&&&&&&&& network & QoE inference \\\hline
\rowcolor{rowcolor1}
{\bf Yang et al.~\cite{yang2021quality}}  & &&&&&&& \cmark & - & Traffic classification \\ \hline
%&&&&&&&&&&& \\\hline
\end{tabular}
\end{center}
\label{tab:sota}
\end{table*}

\subsection*{Development}~\label{sec:dev}
%\textcolor[HTML]{006400}{
Model development is an iterative process. With each cycle, it's important to assess the current model's performance compared to its past versions and determine its readiness for live deployment~\cite{huyen2022designing}. Model development consists of two fundamental steps, i.e., {\em algorithm design,  model training \& validation}, each crucial to determine a solution's overall readiness for the target network.

\textbf{Algorithm design.} The purpose of ML can be threefold: (i) making effective use of \textit{existing} knowledge, (ii) gathering a structured understanding of \textit{unknown} phenomena, and (iii) \textit{learning} to achieve a goal, which can be mapped to three branches, i.e., Supervised, Unsupervised, and Reinforcement Learning (RL) – with potential intersections among them (e.g., semi-supervised or self-supervised learning).

Supervised ML techniques, such as regression and classification, excel at tracking well-specified problems in open-loop settings to increase visibility about network traffic or distill insight from raw data. In particular, regression techniques are fit for forecasting (e.g., traffic demand or user behavior) or learning complex relationships, such as relating network Quality of Service (QoS) indicators to user Quality of Experience (QoE). Classification techniques are another related example where AI techniques are useful: traffic prioritization requires coarse-grained traffic class labels for policing and may additionally require fine-grained application labels.

Unsupervised ML operates by identifying patterns and structures within data without labeling, relying instead on the algorithm's ability to discern intrinsic features and relationships within the dataset. For example, unsupervised AI employs algorithms in anomaly detection to discern data deviations by autonomously learning underlying distributions. These algorithms identify outliers representing significant departures from established patterns without reliance on pre-labeled normal data instances. Lastly, RL is suitable for sustained and efficient closed-loop AI automation environments. An example is the automation of resource management by using RL, implemented through centralized cloud agents or distributed device agents~\cite{rossi2022landing}. In this context, AI agents are dedicated to improving QoS, e.g., enhancing transmission efficiency and reducing latency. To attain such a goal, agents are rewarded for their actions, effectively balancing exploration and exploitation within a vast state space, thus providing automated and optimized solutions~\cite{holland2021new}.

%\textcolor[HTML]{006400}{
%Overall, 
% Choosing models requires navigating through several compromises, including the trade-off between false positives and false negatives (minimizing one may increase the other), the balance between computing needs and accuracy (more sophisticated models might achieve better accuracy but need stronger computing resources), and the trade-off between how interpretable a model is versus its performance.}

%\red{Algorithm design involves selecting the right type of learning algorithm (e.g., SL vs. UL), ML model (e.g., regressions, decision tree, ensembles, DNNs), and architecture (e.g., $\#$layers, $\#$neurons/layer).} 
%The KPIs and network constraints should be jointly contemplated as ML algorithms have divergent predictive powers, resource footprints, and application scenarios. 
%Sometimes, multiple models should be developed and pooled to compensate for the sporadic changes in highly dynamic networks. 
%Similarly, the hyper-parameters should be cognitively adjusted during model tuning to find the optimal configuration. In existing solutions, both processes are typically manually conducted based on domain knowledge, which can be tedious and strenuous for complex models. 
%For network operators, it is preferable to automatize these processes to avoid sub-optimal solutions~\cite{holland2021new}. 

\textbf{Model training \& validation.} 
In the system context of model training \& validation, factors such as inference efficiency, generalizability, and safety hold similar significance as the traditional focus on accuracy. For instance, generalizability ensures timely adaptation in dynamic environments like disaster-resilient networks,  safety is crucial for ML algorithms that require frequent interaction with real systems, and inference efficiency is crucial for quick decision-making. %\textcolor[HTML]{006400}{ 
The process of training and validating models can be enhanced using tools such as MLflow, Weights \& Biases, and DVC. These tools facilitate the selection of ML algorithms and the adjustment of hyperparameters, driving towards automated and efficient optimization of models. 
%, and fairness is essential to avoid biases in critical applications. %Notably, explainability allows for interpretation, auditing, and management of a model's decisions, garnering trust among stakeholders~\cite{zhang2022interpreting}. We further elaborate on these factors in Sec.~\ref{missing}. 

%Although model training and validation are well-studied in existing research, they still miss some key factors once in the system context. For example, a training strategy should factor in efficiency and safety. 
%The former is crucial to delivering up-to-date models in highly dynamic settings like disaster-resilient networks. The latter is necessary for AI/ML algorithms (e.g., reinforcement learning) that call for frequent interactions with real systems. 
%Likewise, inference efficiency, fairness, and explainability during model validation should be examined with the evaluation metric to unveil a model's readiness for the target network. Inference efficiency is critical for real-time analysis and decision-making in high-speed networks (e.g., beyond $10$ Gbps). 
%Fairness is requisite in mission-critical environments to unveil biases and avoid unexpected fallouts. Explainability allows a model's decisions to be interpreted, audited, managed, and ultimately trusted by various stakeholders~\cite{zhang2022interpreting}.
%We further elaborate on these factors in Sec.~\ref{missing}. 
\textit{Existing solutions:} Two prior works explore AutoML to automatically carry out model selection and hyper-parameter tuning to hide the AI/ML-specific complexities from network operators. {Holland et al.~\cite{holland2021new}} leverage the AutoGluon-Tabular framework to locate and ensemble models with high predictive
accuracy and low inference latency, given the features and labels. Similarly, {Swamy et al.~\cite{swamy2022homunculus}} employ an optimization framework that automatically performs algorithm selection and model generation as a Bayesian optimization problem based on user intents and network constraints. {Lacoboaiea et al.~\cite{iacoboaiea2022design}} address the challenges of building a Deep RL-based channel manager, specifically focusing on training safety, efficiency, environment realism, and generalization. They leverage digital twins for secure training, adjust learning rates for efficiency, enhance simulator fidelity with real-world data, and bolster generalization via synthetic noise and actual data integration.

\subsection*{Operations}~\label{sec:ops} 
%Based on our study, most existing solutions neglected the practical challenges of {\em deploying} and {\em managing} AI/ML-based solutions, making them difficult to actuate in real networks. 
This part elaborates on AI/ML-based solutions requiring attention in real networks on {\em deployment} and {\em management}.

\textbf{Deployment:} Operational deployment encompasses packaging, customization, and feasibility tests.
As traditional ML-based solutions were mainly intended for the control plane, which standard model serving tools can handle. Recently, with the rise of in-network ML, researchers began to push the ML frontier into the network data plane to capitalize on the voluminous data there~\cite{zheng2022automating}. Model deployment becomes a Sisyphean task due to the distinctions between the local implementation environment and network infrastructure, and the divergent tooling can sorely impede customization. Moreover, as networks are replete with a plethora of specialized hardware devices (e.g., SmartNICs, P4 switches, embedded devices) 
with disparate architectures, configuration routines, and resource footprints, the deployment process entails refactoring a solution into a generic data-processing pipeline with minimal interference on the network service~\cite{swamy2022homunculus}. %\textcolor[HTML]{006400}{The challenging aspects of deployment involve ensuring the developed model is accessible to millions of users with millisecond latency and near-perfect availability, establishing infrastructure to instantly alert the appropriate individual if an issue arises, diagnosing the problem, and smoothly rolling out updates to rectify the issue.}

\textbf{Management.} Furthermore, managing the deployed ML-based solutions involves model serving, resource \& operation management, and drifting monitoring tasks. % \textcolor[HTML]{006400}{Deploying a model is not the end of the process, and the performance of a model degrades over time in production. } 
In particular, as network systems can evolve expeditiously, the intrinsic concept/data drifts can result in model decay and service degradation. The inference quality should thus be constantly inspected to detect performance diminishments and trigger the model-rebuilding process whenever applicable. In real networks, the correct quality metrics and triggers should be carefully scoped, and the monitoring overhead should also be balanced with the quality assessment accuracy~\cite{yang2021quality}. Depending on the problem context, the rebuilding process can start from the data preparation and labeling or model development stage, which must be specified beforehand.

\textit{Existing solutions.} To cope with these challenges, {Zheng et al.~\cite{zheng2022automating}} introduce the {Planter}, a modular architecture that facilitates the seamless deployment of diverse in-network ML algorithms across three prominent hardware platforms. %: Intel Tofino, BMv2, and P4Pi.
Planter accommodates a slew of mainstream ML algorithms. Its post-training automatically converts the models into tailored P4 code for specific targets, subsequently undergoing compilation and integration for deployment. Similarly, {Swamy et al.~\cite{swamy2022homunculus}} craft compiler tools designed to render target-oriented code for popular data planes autonomously.  %platforms: FPGA, Tofino, and Taurus.
They harness a cycle-accurate simulator to preemptively gauge the model kPIs, encompassing throughput, latency, and resource utilization. {Yang et al.~\cite{yang2021quality}} tackle inference monitoring and combine gradient-based techniques with Open Set Recognition \& explainable AI to scrutinize inference qualities. Comparative evaluations have been conducted to validate the proficiency of their approach in inference monitoring and data drift detection.

%These works constitute vital advancements in deploying and managing pre-trained AI/ML models in real networks. As network vendors introduce more sophisticated features and devices, the supported ML models, deployment targets, and monitoring methods will keep enriching, which the tools and methods of these works can assist with. 

We summarize all these pioneering works in Table~\ref{tab:sota} regarding the tackled lifecycle stages, supported types of ML algorithms, targeted network environment, and use cases. Essentially, each work covers part of the ML lifecycle stages.

\begin{figure}
    \centering
    \includegraphics[width=0.5\textwidth]{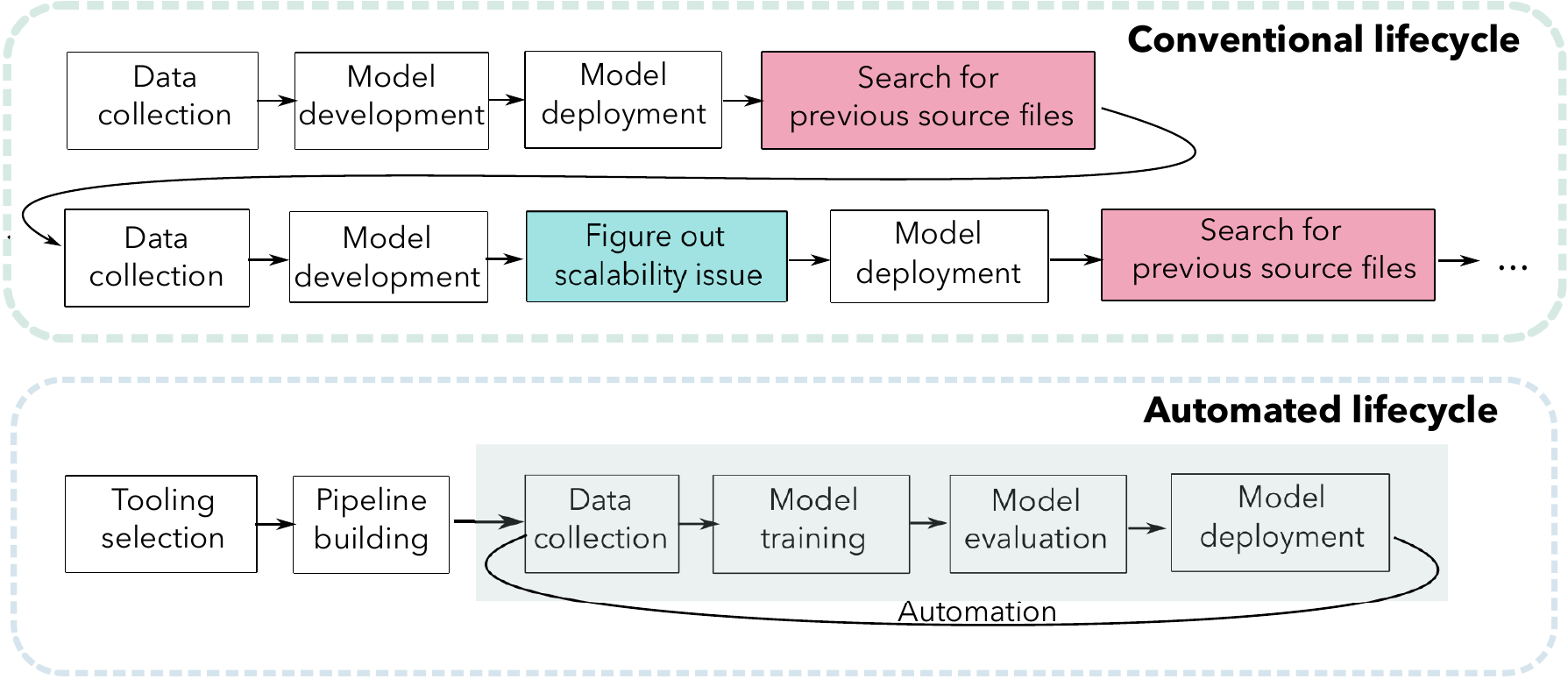}
    \caption{Conventional vs. Automated  ML lifecycle. }
    \label{fig:comparasion}
\end{figure}

\iffalse
***************reviewer 2***************
reviewer2:Continuous learning was correctly characterized as a standing issue leading to the section on MLOps, yet it appears not to have been addressed in that section nor as the paper concludes. Would a remark on how MLOps helps with this be feasible?
***************reviewer 2***************
\fi

\subsection*{Missing pieces to the puzzle}\label{missing}
Based on the proceeding review, we identify three missing pieces to the fully operationalized AI/ML puzzle. 
First, despite the optimistic individual advancements, they have not been cumulatively translated into global benefits. In real systems, individual stages must be seamlessly articulated as an end-to-end data processing pipeline. %to reduce the management \& maintenance overhead.
With the current reliance on manual interventions, ML-based solutions will become heavy to manage in future networks. 
Second, reproducibility is not enforced due to the absence of systematic logging and tracking. Traditional version control tools cannot sufficiently capture the nuances of ML workflows' datasets, parameters, and configuration dependencies, which must be consistently reproducible for scientific rigor and regulatory compliance.
Third, silos can also arise due to the disparate expertise \& priorities of data scientists and network engineers, hampering productivity and stalling time-to-value.

Fig.~\ref{fig:comparasion} illustrates two approaches for ML lifecycle management. The traditional workflow is a one-off process of data collection, model development, and deployment. This approach prioritizes rapid delivery for the initial time. Nonetheless, as the temporal dimension extends, this method becomes less efficient. In particular, the data/system shifts necessitate continuous model retraining. Without proper management, reproducing and enhancing existing models become laborious as the whole process can involve multiple teams, from data scientists to network engineers. Manual asset transferring is inefficient and burdensome.  

Conversely, the second approach adopts a more systematic strategy. Initially, the involved teams dedicate significant time to constructing an automated pipeline with established tracking mechanisms.
Compared to the manual approach, it confers substantial long-term benefits, including reproducibility, continuous model enhancement, and seamless communication.
%Being fully automated, the process ensures consistent reproducibility, enabling continuous model enhancement, and paves the way for concerted efforts across different teams.

\subsection*{Continual learning}\label{Continual}
%Continual learning involves creating a framework that enables data scientists or AI engineers to update models as necessary and deploy them quickly, including building from scratch and fine-tuning. It's important to differentiate between ``continual learning" and ``continuous learning"; the latter refers to models that learn from each incoming sample, while the former involves learning in batches or micro-batches~\cite{huyen2022designing}. The common application of continual learning can address the data distribution drift issue, adjust models based on rare events, and solve the cold start problem, which arises when a model makes predictions for a new scenario without any historical data. Focusing on the context of network systems, we list the progression toward implementing continual learning below:

Continual learning enables AI/ML practitioners to update and deploy models efficiently. 
%It's important to differentiate between ``continual learning" and ``continuous learning"; the latter refers to models that learn from each incoming sample, while the former involves learning in batches or micro-batches~\cite{huyen2022designing}. 
It addresses data distribution drifts, adjusts models based on rare events, and solves the cold start problem arising from unseen data~\cite{huyen2022designing}. With respect to network systems, we enumerate the progression toward continual learning below:

{\bf Stage 1 - Manual, stateless retraining:} %Initially, many researchers, particularly those who recently integrated ML into network operations, manually retrain their models without leveraging historical data state. This approach is common in settings without dedicated teams to manage ML platforms.
Initially, researchers manually retrain models without leveraging historical data state, which is common in settings without dedicated teams to manage ML platforms.

{\bf Stage 2 - Automated retraining:} 
%At this stage, researchers begin automating their network models' retraining process. The frequency of retraining often relies on intuition, such as daily updates during periods of low network usage, aiming to optimize performance without a solid empirical basis.
Researchers begin automating model retraining. The retraining frequency often relies on intuition, such as daily updates, to optimize performance without a solid empirical basis.

{\bf Stage 3 - Automated, stateful training:} 
%To enhance efficiency, this stage introduces improvements to the retraining process by starting from the latest saved state of the network model. This method particularly benefits networks requiring frequent updates, reducing the computational overhead by leveraging previous training checkpoints.
To improve efficiency, researchers begin exploring the recently saved model states and checkpoints, which is especially beneficial for use cases requiring frequent model updates. 

{\bf Stage 4 - Continual learning for network management:}
%The most advanced stage involves transitioning from fixed-schedule updates to dynamic, trigger-based model updates. These triggers might be based on time intervals, performance metrics, network volume changes, or traffic pattern shifts, enabling more responsive and adaptive network management.
The most advanced phase involves transitioning from fixed-schedule updates to dynamic, trigger-based model updates based on time intervals, performance metrics, network volume, or traffic patterns, enabling more responsive and adaptive network management.

%Applying continual learning in modern networks faces significant hurdles, notably in accessing fresh data and evaluating models. Despite these challenges, the maturation of MLOps tools for continual learning suggests a future where implementing it could become as straightforward as batch learning. We will introduce MLOps in the next section. 
Applying continual learning in modern networks faces significant hurdles. Fortunately, MLOps provides means to offset them, as detailed in the next section.

\section*{MLOps: towards end-to-end pipelines} 
\label{sec:future}

%\textcolor[HTML]{006400}{
%Ops in MLOps from DevOps, short for Developments and Operations. To operationalize something means to bring it into production,  which includes deploying, monitoring, and maintaining it. MLOps is a set of tools for bringing ML into production.%}

MLOps is an emerging set of practices that apply DevOps principles to unify the development and operation of ML-based systems~\cite{databricks,huyen2022designing}.

\subsection*{Why MLOps?}\label{sec:mlops} 

\iffalse
*********************reviewer1**************
\footnote{\textcolor{red}{Reviewer 1 wants proof: The authors claim that 'MLOps is an emerging set of engineering practices in the ML field aiming at applying DevOps principles to unify the development and operation of ML-based systems.' How? What work can prove this? How this is done?. The present work, however, fails to prove the author's point.}}
*********************reviewer1**************
\fi
Traditionally, the operational costs of delivering software products can be countered with DevOps, which encompasses an assemblage of principles to break the silo between software developers and IT operations engineers, promoting Automation and Continuous Integration (CI) / Continuous Deployment (CD) throughout the product lifecycle.
These principles help drive IT and business outcomes for many businesses and organizations~\cite{yang2021quality}. The network community has adopted DevOps to fuel technological innovation and revenue growth. 

However, though DevOps can curb the operational overhead of productionalizing traditional software projects, they lack supplemental support for the unique characteristics of ML. There are five fundamental discrepancies between conventional software and ML:  First, code quality predominantly decides the performance in traditional software; In AI/ML, the model and data all impact the outcome~\cite{databricks}. Second, traditional software is usually built on full-fledged libraries with clear abstraction boundaries~\cite{sculley2015hidden}. ML-based solutions often involve a broader range of tools and libraries, subject to extra integration and maintenance costs. Third, unlike traditional software that conveys deterministic outputs, ML models are intrinsically stochastic and entail disparate processes to validate their behaviors. Fourth, ML models are susceptible to data/concept drifts, which are common in real networks and thus necessitate drift detection and model rebuilding~\cite{yang2021quality}. Finally, building and operating ML-based solutions call for data science skillsets, which are missing in traditional software/network routines. According to a recent survey, $55\%$ of telcos lack the pertinent data science talent~\cite{ava}. %Although network practitioners can gradually get acquainted with AI/ML and data science, mastering the theoretical and practical aspects requires time. 

\begin{figure}[!tb]
\begin{center}
\includegraphics[width=0.49\textwidth]{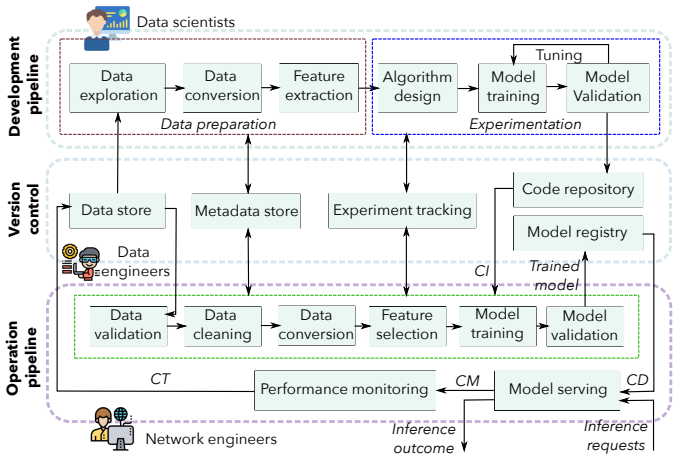}
\caption{Operationalizing AI/ML in  Future Networks: A Bird's Eye View from the System Perspective}
\label{fig:mlops}
\end{center}
\end{figure}

%\footnote{\textcolor{blue}{reviewer 2: The caption for Figure 3 might as well be the paper's title.}}
%\subsection*{What is MLOps?} 
% A collection of specialized practices and tools have been proposed to reshape the DevOps principles for ML-based products since 2015, forming the MLOps discipline. 
Layered on the DevOps tenets, MLOps accommodates the unique traits of AI/ML with the following practices: 

\begin{itemize}
\item {Continual Monitoring (CM) / Continual Training (CT)}: MLOps addresses the model decay problem by constantly monitoring the data and inference quality and rebuilding the model whenever applicable. 
\item {Automation}: MLOps streamlines AI/ML lifecycle into a fully automated pipeline to alleviate operational costs. %without human intervention. 
%\item {\bf Testing}: Besides unit and integration tests, MLOps consolidates data, feature, and model testing to fully ascertain a solution's compatibility with the whole network system. 
\item {Versioning}: Based on DevOps, MLOps extends the version control of artifacts involved in the process, including data, model, and code. %The accompanying data and feature stores also simplify data governance. 
\item {{Experiment tracking}}: Experiments are systematically tracked to ensure reproducibility and auditability. 
\item {{Collaboration}}: MLOps advocates a common platform to build synergy across the involved participants. %with different priorities and expertise.
\end{itemize}

With these practices, MLOps consolidates innovations across the AI/ML lifecycle and dramatically curtails operational costs, even though this burgeoning discipline is still nascent for the network research community. We envision a plausible architecture in Fig.~\ref{fig:mlops}, which adopts most MLOps practices for real networks. %, though considerable tooling and engineering efforts are still inevitable. 

\subsection*{MLOps for networking: A Case study}
\begin{figure}[!tb]
    \centering
    \includegraphics[width=\linewidth]{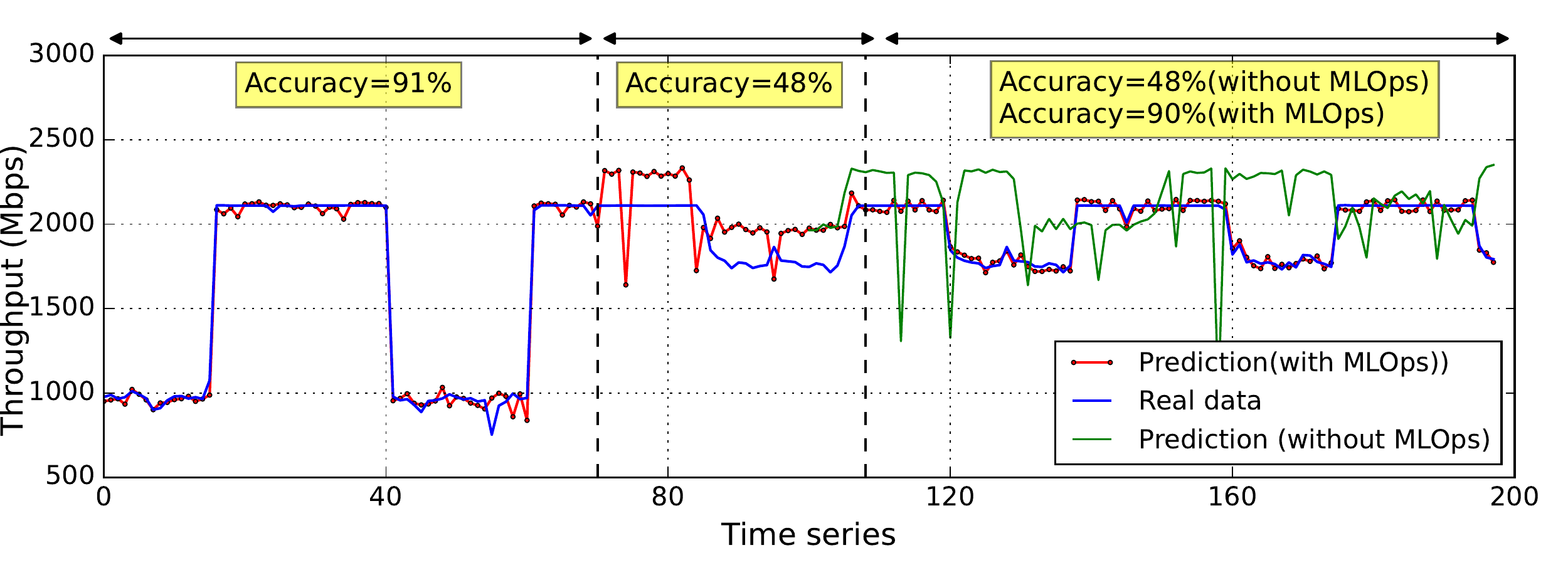}
    \caption{The benefits of MLOps for KPI prediction}
    \label{fig:mlops_illustration_1}
\end{figure}

We demonstrate the advantages of MLOps through a case study on real-time KPI prediction, a critical aspect of network management. %We deploy a network service chain across a distributed cluster inside a small-scale enterprise data center; each node is a commodity server equipped with 10/40-Gigabit network cards. 
%\subsubsection*{KPI prediction} We consider non-intrusive KPI prediction, which is an actively investigated topic~\cite{manousis2020contention}. Instead of in-band data collection, which leads to non-trivial engineer exertion and data-path overhead, %we continually monitor hardware subsystems and prediction the end-to-end service throughput. 
We deploy a network service chain in a small-scale data center and explore a lightweight ANN model for ``non-intrusive" KPI prediction using infrastructural-level hardware features. We employ the Pearson Correlation Coefficient for feature selection, Bayesian optimization for automatic hyperparameter tuning, and Jensen-Shannon divergence to quantify data drift. We restructured the processing pipeline using Kubeflow, an open-source MLOps platform based on Kubernetes. Fig.~\ref{fig:mlops_illustration_1} demonstrates how MLOps enables real-time KPI prediction with sustainable performance. Initially, our model achieved an average prediction accuracy of $91\%$ for service throughput. At timeslot 70, a data drift occurred, degrading the accuracy to $48\%$. In response, the system automatically triggered model retraining, which rolled out an updated model that restored the prediction accuracy to $90\%$.

\section*{Conclusion}\label{sec:conclusion} 
Due to the lack of system-related considerations, AI/ML is still not an integral part of modern networks. This paper analyzed the inconsistencies between existing AI/ML-based solutions and real network systems and discussed all the practical considerations throughout their product lifecycle. We also reviewed the related works and identified the missing pieces. Then, we conducted a case study to validate the advantages of MLOps in a real network system.
%Based on our experience, we recommend MLOps as a promising way to erase operational concerns. 
This paper can raise awareness about the practical hurdles of operationalizing AI/ML in production settings and expedite its integration into future networks.

\bibliographystyle{IEEEtran}
\bibliography{IEEEabrv,biblio}

\begin{IEEEbiographynophoto}{Qiong Liu}  (Member, IEEE)
    is a postdoc researcher at Telecom Paris. She received her B.S. degree from Shandong University, China 2015. She received the M.S. degree from Xidian University in 2018, and the Ph.D degree from INSA Rennes, France in 2022. Currently, she focuses on applied AI for network systems and stochastic geometry-based performance evaluation in large-scale networks.
\end{IEEEbiographynophoto}
    
\begin{IEEEbiographynophoto}{Tianzhu Zhang}(Member, IEEE)
is a research scientist at Nokia Bell Labs. He received his B.S. degree from Huazhong University of Science and Technology, China, in 2012. He received his M.S. and Ph.D. degrees in 2014 and 2017 from Politecnico di Torino, Italy. From 2017 to 2019, he was a PostDoc researcher at Telecom ParisTech. His research interests center around applied AI/ML for network systems.
\end{IEEEbiographynophoto}

\begin{IEEEbiographynophoto}{Masoud Hemmaptour}
received the MS degree in computer and communication network engineering and the PhD degree in control and computer engineering from Politecnico di Torino, Italy, in 2015, and 2019, respectively. His research interests include high-performance interconnect and programmable network devices, Currently, he focuses on the performance and energy efficiency of in-network processing applications.
\end{IEEEbiographynophoto}

\begin{IEEEbiographynophoto}{Han Qiu}
received the B.E. degree from the Beijing University of Posts and Telecommunications, China, in 2011, the M.S. degree from Institute Eurecom, France, in 2013, and the Ph.D. degree from the Department of Networks and Computer Science, Telecom-ParisTech, France, in 2017. He worked as a postdoc at Telecom Paris from 2017 to 2020. He is an assistant professor at the Institute for Network Sciences and Cyberspace, Tsinghua University, China. His research interests include AI \& Data security and cloud
computing.
\end{IEEEbiographynophoto}

\begin{IEEEbiographynophoto}{Dong Zhang} (Member, IEEE) received the B.S. and Ph.D. degrees from Zhejiang University, China, in 2005 and 2010, respectively. He visited Alabama University, USA, as a Visiting Scholar from 2018 to 2019. He is currently a Professor at the College of Computer Science and Big Data, Fuzhou University, China. His research interests include software-defined networking, network virtualization, and Internet QoS.
\end{IEEEbiographynophoto}

\begin{IEEEbiographynophoto}{Chung Shue Chen}(Senior Member, IEEE)
received the B.Eng., M.Phil., and Ph.D. degrees in information engineering from the Chinese University of Hong Kong (CUHK), Hong Kong, in 1999, 2001, and 2005. He is a DMTS at Nokia Bell Labs. %Prior to joining Bell Labs, he worked at INRIA in the research group on Network Theory and Communications (TREC, INRIA-ENS). He was an Assistant Professor at CUHK. He was an ERCIM Alain Bensoussan Fellow with the Norwegian University of Science and Technology (NTNU), Norway, and the National Centre for Mathematics and Computer Science (CWI), The Netherlands. He worked at CNRS in Lorraine on Real-Time and Embedded Systems. 
His research interests include wireless networks, communications, optimization, machine learning, 5G/6G, IoT, and intelligent systems. %Dr. Chen received the Sir Edward Youde Memorial Fellowship and the ERCIM Fellowship. He was a TPC in international conferences, including IEEE ICC, Globecom, WCNC, PIMRC, VTC, CCNC, and WiOpt (a TPC Vice Chair). He is an Editor of the Transactions on Emerging Telecommunications Technologies (ETT). He is a Permanent Member of the Laboratory of Information, Networking and Communication Sciences (LINCS), France. 
\end{IEEEbiographynophoto}

\begin{IEEEbiographynophoto}{Marco Mellia} (M’97–SM’08-F'20)
%is a full professor at the Control and Computer Engineering Department of Politecnico di Torino. In 2002, he visited the Sprint Advanced Technology Laboratories, working at the IP Monitoring Project (IPMON). From 2011 to 2013, he collaborated with Narus Inc, CA, working on traffic monitoring and cybersecurity system design. In 2015 and 2016, he collaborated with Cisco Systems to design cloud monitoring platforms based on machine learning.  He is now the coordinator of the SmartData@PoliTO Center on Data Science and Machine Learning, involving more than 50 colleagues and Ph.D. students. Prof. Mellia co-authored over 250 papers published in international journals and presented at leading conferences, all of them in communication networks. He won the IRTF ANR Prize at IETF-88 and the best paper award at IEEE P2P’12, ACM CoNEXT’13, and IEEE ICDCS’15. He is the Area Editor of ACM CCR, IEEE Transactions on Network and Service Management, and Elsevier Computer Networks.  His research interests are in the area of traffic monitoring and big data analysis, with applications to traffic classification, management, and security. 
is a full professor at the Control and Computer Engineering Department of Politecnico di Torino and the coordinator of the SmartData@PoliTO Center on Data Science and Machine Learning. 
%He co-authored over 250 papers published in international journals and presented at leading conferences, all of them in communication networks. %He won the IRTF ANR Prize at IETF-88 and the best paper award at IEEE P2P’12, ACM CoNEXT’13, and IEEE ICDCS’15. He is the Area Editor of ACM CCR, IEEE Transactions on Network and Service Management, and Elsevier Computer Networks.  
His research interests include traffic monitoring and big data analysis, with applications to traffic classification, management, and security. 
\end{IEEEbiographynophoto}

\begin{IEEEbiographynophoto}{Armen Aghasaryan}
holds a PhD in signal processing and telecommunications from INRIA / University of Rennes, France. He joined Alcatel in 2000 and is heading the Machine Learning \& Systems group in the AI Research Lab, Nokia Bell Labs. His research interests include AI/ML, Data Analytics, Cloud Computing, and Network Automation. 
\end{IEEEbiographynophoto}

\end{document}